\documentclass[twocolumn,english,showpacs]{revtex4}
\usepackage{babel,amsmath,amssymb,dcolumn}
\usepackage[dvips]{graphics}
\usepackage[dvips]{graphics}
\usepackage{amsmath}
\usepackage{amssymb}
\usepackage{latexsym}
\usepackage{graphicx}
\usepackage{psfrag,fancyhdr,epsfig}
\newcommand{\bea}{\begin{eqnarray}}
\newcommand{\eea}{\end{eqnarray}}
\newcommand{\bs}{\begin{slide}}
\newcommand{\es}{\end{slide}}
\newcommand{\bi}{\begin{itemize}}
\newcommand{\ei}{\end{itemize}}
\newcommand{\beq}{\begin{equation}}
\newcommand{\eeq}{\end{equation}}
\newcommand{\om}{\omega}

\begin{document}

\title{Quasi-Normal Modes of Brane-Localised Standard Model Fields II: Kerr Black Holes}
\author{P. Kanti}

\email{panagiota.kanti@durham.ac.uk}

\affiliation{Department of Mathematical Sciences, University of Durham\\
Science Site, South Road, Durham DH1 3LE, United Kingdom}

\author{R. A. Konoplya}

\email{konoplya@fma.if.usp.br}

\affiliation{Instituto de F\'{\i}sica, Universidade de S\~{a}o Paulo \\
C.P. 66318, 05315-970, S\~{a}o Paulo-SP, Brazil}

\author{A. Zhidenko}\email{zhidenko@fma.if.usp.br}

\affiliation{Instituto de F\'{\i}sica, Universidade de S\~{a}o Paulo \\
C.P. 66318, 05315-970, S\~{a}o Paulo-SP, Brazil}


%

\begin{abstract}
This paper presents a comprehensive study of the fundamental quasinormal
modes of all Standard Model fields propagating on a brane embedded in a
higher-dimensional rotating black hole spacetime. The equations of motion
for fields with spin $s=0, 1/2$ and 1 propagating in the induced-on-the-brane
background are solved numerically, and the dependence of their QN spectra
on the black hole angular momentum and dimensionality of spacetime is
investigated. It is found that the brane-localised field perturbations
are longer-lived when the higher-dimensional black hole rotates faster,
while an increase in the number of transverse-to-the-brane dimensions
reduces their lifetime. Finally, the quality factor $Q$, that determines
the best oscillator among the different field perturbations, is investigated
and found to depend on properties of both the particular field studied
(spin, multipole numbers) and the gravitational background (dimensionality,
black hole angular momentum parameter).

\end{abstract}

\maketitle


\section{Introduction}

The introduction, during the recent years, of gravitational theories that
postulate the existence of additional spacelike dimensions in nature
\cite{RS, ADD} have radically changed our notion of the universe and of the
structure of the spacetime in which we live. The concept of the brane -- a
(3+1)-dimensional hypersurface inspired by D-branes encountered in
string theory -- was introduced, that was embedded in a higher-dimensional
spacetime, the bulk. Ordinary Standard Model (SM) particles (scalars, fermions
and gauge bosons) are restricted to live on  the brane while gravitons
can propagate both on and off the brane. In this set-up, the accurately
observed properties of SM fields are only marginally altered, nevertheless
new effects coming from the embedding of our 4-dimensional world in a
higher-dimensional spacetime could still be observed, despite the fact that
the bulk itself remains strictly non-accessible to brane observers.

One such effect could be the creation of microscopic black holes on ground-based
accelerators during the collision of highly energetic particles with center-of-mass
energy $\sqrt{s}> M_*$ \cite{creation}, where $M_*$ denotes the fundamental
Planck scale of the higher-dimensional gravitational theory. This scenario
can be realized in the context of the theory with Large Extra Dimensions
\cite{ADD} that allows for a fundamental gravity scale as low as 1 TeV.
As a result, experiments performed at next-generation particle colliders
could easily witness effects coming from trans-planckian collisions, such
as the creation of microscopic, higher-dimensional black holes that
are centered on our brane but extend off the brane as well. The creation
of these black holes, and thus the existence of additional dimensions,
could be confirmed through the detection of the emitted Hawking radiation
which is expected to be their most prominent observable effect \cite{Kanti}.

Nevertheless, the creation of these brane black holes might trigger additional
observable effects, such as the detection of the spectrum of the so-called
quasi-normal modes (QNMs) \cite{Nollert1, KS}. These modes arise as the result
of an external perturbation of a black hole background, either through the
addition of a field or by perturbing the metric itself. When this happens,
the gravitational system enters a phase of damping oscillations, with the
frequency of the field consisting of a real part $\omega_{\rm Re}$, that
drives the field oscillations, and of an imaginary part $\omega_{\rm Im}$,
that cau\-ses the simultaneous damping of these oscil\-lations. Quasi-normal
modes with small $\omega_{\rm Im}$, and thus long damping time, can dominate
the spectrum at very late times after the initial perturbation.

As the detection of the spectrum of QNMs could be considered as direct
evidence for the existence of black holes in our universe, a variety of
black-hole backgrounds and auxiliary propagating fields have been studied
over the years both in 4 dimensions \cite{quasi-4D} and higher dimensions
\cite{quasi-D}. Unfortunately, no experimental detection has so far
confirmed these theoretical studies. This could be due either to the
notorious elusiveness of gravitons or to the fact that simply no black hole
exists, as yet, in the vicinity of our neighbourhood. If, however, the scenario
of the creation of microscopic black holes can be realised at the next-generation
colliders, a unique opportunity for detecting the QN spectrum is offered.
Since we have not yet successfully detected any gravitons, the obvious
candidates would be instead the different species of SM fields that are
not only abundant on our brane but also much easier to detect.

In \cite{KK}, we presented a comprehensive study of the QN spectra
of all SM fields (scalars, fermions and gauge bosons) living on a
4-dimensional brane, that was embedded in a variety of
higher-dimensional, spherically-symmetric black-hole backgrounds.
The cases of brane-localised fields propagating in a
Schwarzschild, Reis\-sner-Nordstr\"om and Schwarzschild-(Anti) de
Sitter induced gravitational background were studied in detail,
and the effect of various parameters, such as the dimensionality
of spacetime, black hole charge and bulk cosmological constant, on
the QN spectra was examined. It was found that an increase in the
number of transverse-to-the-brane spacelike dimensions resulted in
the faster damping of all fields living on the brane,
independently of the exact type of spherically-symmetric projected
background. In the presence of a black hole charge, the QN spectra
of all fields resembled more the ones of $D$-dimensional fields
rather than the ones of 4-dimensional, while the presence of a
bulk cosmological constant did not lead to significant
modifications to the QN spectra compared to the ones of purely
4-dimensional fields. A recent paper \cite{elcio} follows the same
line of research by studying the quasinormal modes of
brane-localised black holes and their thermodynamic properties.

In the present work, we extend our previous analysis by considering
higher-dimensional black-hole backgrounds with non-vanishing angular
momentum. During the high-energy collisions of elementary particles
resulting in the creation of black holes, it is unnatural to expect that
only head-on collisions, leading to spherically-symmetric black holes,
would take place. Collisions with a non-vanishing impact parameter are
much more likely to occur, and, in addition, it is for these collisions
that the black-hole production cross-section is maximised \cite{creation,
Kanti}. Therefore, microscopic rotating black holes should be the most
generic situation, and the effect of the angular momentum of the black
hole on the QN spectra of brane-localised fields needs to be investigated.
An earlier paper \cite{Berti} that appeared in the literature studied the
QN spectra of brane-localised SM fields only for the particular
case of a 5-dimensional rotating black hole. Here, we extend this
work by presenting a comprehensive analysis of the QN spectra of fields
with spin $s=0,1/2$ and 1 for arbitrary dimensionality of spacetime and
angular momentum of the black hole.

As in our previous analysis, in order to ignore quantum corrections, we
will be assuming that the produced black hole has a mass $M_{BH}$ that
is at least a few orders of magnitude larger than the fundamental scale
of gravity $M_*$. Also the brane self-energy can be naturally (i.e. in
the context of a theory that solves the hierarchy problem) assumed to be
of the order of $M_*$; therefore, it is much smaller than $M_{BH}$,
and its effect on the gravitational background can also be ignored.

In section II, we present the theoretical framework for our analysis and
the equations of motion for the SM fields propagating in the brane background.
Section III outlines the numerical techniques employed for the integration
of both the radial and angular part of the equation of motion of a field
propagating in an axially-symmetric induced background. Our numerical
results are presented in Section IV, where the effect of the dimensionality
of spacetime, angular momentum of the black hole and spin of the particle
on the QN spectra is investigated. We finish with your conclusions in Section V.


\section{Master equation for propagation of fields on the brane}

The line-element describing a higher-dimensional, rotating, neutral
black hole is given by the Myers-Perry solution \cite{Myers}. As mentioned
above, in this work we will concentrate on the propagation of SM fields,
i.e. fields with spin $s=0$, 1/2 and 1, on the induced-on-the-brane
gravitational background. This background is given by the projection of
the higher-dimensional one onto the brane by fixing the values of the
additional angular coordinates that describe the $(D-4)$ extra spacelike
dimensions \cite{kmr1, kmr2}. After the projection, the brane background
assumes the form \cite{Kanti}
\begin{equation}
\begin{split}
ds^2=\left(1-\frac{\mu}{\Sigma\,r^{D-5}}\right)dt^2&+\frac{2 a\mu\sin^2\theta}
{\Sigma\,r^{D-5}}\,dt\,d\varphi-\frac{\Sigma}{\Delta}dr^2 \\[3mm] &\hspace*{-3cm}
-\Sigma\,d\theta^2-\left(r^2+a^2+\frac{a^2\mu\sin^2\theta}{\Sigma\,r^{D-5}}\right)
\sin^2\theta\,d\varphi^2,
\end{split} \label{brane}
\end{equation}
where
\begin{equation}
\Delta=r^2+a^2-\frac{\mu}{r^{D-5}} \quad\mbox{ and } \quad\Sigma=r^2+a^2\cos^2\theta\,.
\label{master}
\end{equation}
The parameters $\mu$ and $a$ are related to the mass and angular momentum,
respectively, of the black hole through the definitions \cite{Myers}
\begin{equation}
M_\text{BH}=\frac{(D-2)\,\pi^{(D-1)/2}}{\kappa^2_D\,\Gamma[(D-1)/2]}\,\mu\,, \qquad
J=\frac{2}{D-2}\,M_\text{BH}\,a\,, \label{parameters}
\end{equation}
with $\kappa^2_D=8\pi G=8 \pi/M_*^{D-2}$ the $D$-dimensional Newton's constant.
We should note here that the higher-dimensional black hole is assumed to have
only one non-vanishing component of angular momentum, about an axis in the brane.
This is due to the simplifying assumption that the particles that created the
black hole were restricted to live on an infinitely-thin brane \cite{ADD},
therefore, during collision they had a non-vanishing impact parameter only on
a 2-dimensional plane along our brane. Also, one should
observe that, as in the case of spherically-symmetric black holes, the
induced-on-the-brane background has an explicit dependence on the total
number of dimensions including the ones transverse to the brane \cite{Kanti}.

For the study of the quasi-normal modes of all SM fields living on the
brane, we need first to derive their equations of motion for propagation
in the induced background (\ref{brane}). For this, the following factorised
ansatz
\beq
\Psi_s(t,r,\theta,\varphi)=  e^{-i\om t}\,e^{i m \varphi}\,\Delta^{-s}\,
P_s(r)\,S_{s,\ell}^m(\theta)\,,
\eeq
is employed for a field with spin $s$, where $S_{s,\ell}^m(\theta)$ are the
spin-weighted spheroidal harmonics \cite{Goldberg}. Then, the use of the
Newman-Penrose \cite{NP, Chandrasekhar} formalism may lead to a ``master"
equation describing the propagation of all species of SM fields (scalars,
fermions and gauge bosons) on the brane. This equation has been derived
in \cite{Kanti,IOP1}, and, as in the case of a purely 4-dimensional Kerr
background \cite{Teukolsky}, it leads to two decoupled equations for the radial
\bea
&~& \hspace*{-0.5cm}
\Delta^{s}\,\frac{d \,}{dr}\,\biggl(\Delta^{1-s}\,\frac{d P_s}{dr}\,\biggr)
\nonumber \\[2mm] &~& \hspace*{0.5cm} + \,\,
\biggl(\frac{K^2-isK \Delta'}{\Delta} + 4i s\,\om\,r
- \tilde \lambda \biggr)\,P_s=0\,,
\label{radial}
\eea
and angular part
\bea
&~& \hspace*{-0.5cm}
\frac{1}{\sin\theta}\,\frac{d \,}{d \theta}\,\biggl(\sin\theta\,
\frac{d S_{s,\ell}^m}{d \theta}\biggr) + \biggl[-\frac{2 m s \cot\theta}
{\sin\theta} - \frac{m^2}{\sin^2\theta} + \label{angular}\\[3mm]
&~& \hspace*{-0.5cm}
+\, a^2 \omega^2 \cos^2\theta - 2 a \omega s \cos\theta +s -
s^2 \cot^2\theta + \lambda \biggr]\,S_{s,\ell}^m=0\,, \nonumber
\eea
of the field, respectively. In the above, we have used the definitions
\beq
K=(r^2+a^2)\,\omega - a m\,, \qquad
\tilde \lambda=\lambda + 2s + a^2 \omega^2 - 2 a m \omega\,.
\label{defs}
\eeq
As in 4 dimensions \cite{Teukolsky}, the above equations hold for the
upper component $s=|s|$ of all SM fields with spin $s=0$, 1/2 and 1,
but in this case they describe their propagation not in a purely
4-dimensional background but on a brane embedded in a higher-dimensional
Kerr background.


\subsection{Alternative form: One-Dimensional wave equation}

As in the case of the study of QNMs of SM fields living on a projected-on-the-brane
spherically-symmetric black-hole background \cite{KK}, we will find it convenient to
rewrite the radial equation (\ref{radial}) in the form of a one-dimensional
Schr\"odinger (or, wave-like) equation. Here, we generalise both the analysis
of \cite{Khanal} performed for the case of a purely 4-dimensional Kerr-like
background, as well as the one of \cite{KK} where a $D$-dimensional
spherically-symmetric background was projected on the brane. To this end,
we define a new radial function and a new (``tortoise'') coordinate according to
\beq
P_s=\rho^{2(s-1/2)}\,Y_s\,, \quad \qquad
\frac{dr_*}{dr}=\frac{\rho^2}{\Delta}\,,
\eeq
where
\beq
\rho^2= r^2 + a^2 -am/\omega \equiv r^2 + \alpha^2\,,
\eeq
and the metric function $\Delta(r)$ is given in Eq. (\ref{master}). Then,
Eq. (\ref{radial}) takes the form
\beq
\left(\frac{d^2}{dr_*^2} + \omega^2\right) Y_s + P\,
\left(\frac{d}{dr_*} + i\omega\right) Y_s - Q\,Y_s=0\,,
\label{new}
\eeq
where we have defined
\beq
P=s \left(\frac{4 r \Delta}{\rho^4} - \frac{\Delta'}{\rho^2}\right)\,,
\eeq
and
\bea
&~& \hspace*{-1.0cm}
Q=\frac{\Delta}{\rho^4}\,\left\{\tilde\lambda - (2s-1)\,\left[
\frac{\Delta}{\rho^2}\right. \right. + \nonumber \\[2mm]
&~& \hspace*{1cm}\left. \left. +\,(2s-3)\,\frac{\Delta r^2}{\rho^4}-
(s-1)\,\frac{\Delta' r}{\rho^2}\right]\right\}.\label{Q}
\eea
By defining further, as in \cite{Khanal},
\beq
Y_s=h Z_s + 2i\om \left(\frac{d}{d r_*} - i \om\right)Z_s\,,
\eeq
Eq. (\ref{new}) can now be written as a one-dimensional Schr\"o\-dinger equation
\beq
\left(\frac{d^2}{dr_*^2} + \omega^2\right) Z_s = V_s\,Z_s\,,
\label{wave}
\eeq
with the effective potential $V_s$ given by
\beq
V_{s=1}=\frac{\Delta}{\rho^4}\,\left[\tilde\lambda - \alpha^2
\frac{\Delta}{\rho^4} \mp i \alpha \rho^2\,\frac{d}{dr}\,\left(
\frac{\Delta}{\rho^4}\right)\right],
\label{spin-1-rot}
\eeq
for spin-1 particles,
\beq
V_{s=1/2}=\frac{\Delta}{\rho^4}\,\left[\tilde\lambda \mp \rho^2\,
\frac{d}{dr}\,\left(\frac{\sqrt{\tilde\lambda \Delta}}{\rho^2}\right)\right],
\label{spin-1/2-rot}
\eeq
for spin-1/2 particles, and
\beq
V_{s=0}=\frac{\Delta}{\rho^4}\,\left[\tilde\lambda +
\frac{\Delta}{\rho^2} -\frac{3 \Delta r^2}{\rho^4} +
\frac{\Delta' r}{\rho^2}\right],
\label{spin-0-rot}
\eeq
for spin-0 particles. The QN modes for gauge bosons, fermions and scalars
living on the brane in the vicinity of a Kerr-like induced background can be
found by solving Eq. (\ref{wave}) upon choosing the appropriate effective potential.
Here, we will undertake this task and compute the QNMs of brane localised SM
fields through numerical analysis -- the details of it are explained in the
next section and our results are presented in the following one.


\section{Numerical Analysis}

In order to calculate the QNMs of brane-localized fields, we need to solve
the radial equation (\ref{radial}), or its equivalent form (\ref{wave}).
To this end, we need the expression of the angular eigenvalue $\lambda$,
that appears in the radial equation. For a vanishing black hole angular
momentum, i.e. $a=0$, the angular eigenvalue is simply given by
$\lambda=\ell(\ell+1)-s(s+1)$, where $\ell=s, s+1, s+2\ldots$ are (half)integer
numbers; therefore, the angular equation can be altogether ignored for
the purpose of calculating QN modes. However, in the case of an
axially-symmetric black-hole background, the angular eigenvalue cannot
be written in a closed form, and its exact value can be found only
numerically \cite{Suzuki:1998vy} by using the angular equation
(\ref{angular}).

For $a\neq0$, the spectrum of $\lambda$ for a given value of $\omega$
is a set of complex numbers that cannot be easily put into correspondence
with a set of angular numbers $l$. In order to find the exact value
of $\lambda(\omega)$, and at the same time the solution to Eqs.
(\ref{radial}, \ref{angular}), we follow a technique inspired by
perturbation theory: we start with the known solution of the angular
equation for $a=0$ (i.e. with the spin-weighted spherical harmonics),
and increase $a$ by a small amount; then, for a given energy $\omega$,
we search for the solution of Eqs. (\ref{radial}, \ref{angular})
that is regular and closest to the original one; this solution is
characterized by a unique value $\lambda(\omega)$. By repeating the
same analysis for different values of ($a, \omega$), the spectrum of
$\lambda(\omega)$, as well as the solution to the radial and angular
equations, can be found for a wide range of the angular momentum
parameter $a$.

The quasinormal modes follow from the numerical integration of the
radial equation (\ref{radial}) after imposing the following
boundary conditions \cite{Nollert1, KS}
\begin{eqnarray}
&P_s &\simeq C_+ \exp(i\omega r)/r\,, \quad {\rm as} \quad r \rightarrow \infty\,,
\label{inftyBC} \\[3mm]
&P_s & \simeq C_- (r-r_h)^{-i\kappa}, \quad {\rm as} \quad r \rightarrow r_h\,,
\label{horizonBC}
\end{eqnarray}
where
\beq
\kappa=\frac{\omega r_h (r_h^2+a^2)-mar_h}{(n-1)(r_h^2+a^2)+2r_h^2}\,,
\eeq
and thus correspond to outgoing waves at spatial infinity and to ingoing
waves at the event horizon $r_h$. The black hole horizon is derived by
solving the equation $\Delta(r)=0$.  Unlike the 4D Kerr black hole
for which there is an inner and outer solution for $r_{h}$, for $D\geq5$
there is only one real, positive solution to this equation, satisfying the
relation $r_{h}^{D-3}=\mu/(1+a_*^2)$, where we have defined $a_*=a/r_{h}$.

The radial equation (\ref{radial}) can be solved by using the continued
fraction method \cite{Leaver:1985ax}. For this, the remaining regular singular
points of Eq. (\ref{radial}) need to be identified. For $D=5$, there is only
one more singularity at $r=-r_h$. That is why the Frobenius series has a
form similar to the one of the ordinary 4-dimensional Kerr background
\begin{equation}
P_s(r)=\frac{\exp(i\omega r)}{r+r_h}\left(\frac{r-r_h}{r+r_h}\right)^{-i\kappa}
\,\sum_{i=0}^\infty b_i\left(\frac{r-r_h}{r+r_h}\right)^i.
\end{equation}
For $D>5$, there are $D-3$ additional singularities, one of them
being at $r=0$. Fortunately, for $D\leq9$, $|1-r_h/r|>1$ for all of
them, thus the appropriate Frobenius series in this case has the
following form
\begin{equation}
P_s(r)=\frac{\exp(i\omega r)}{r}\left(1-\frac{r_h}{r}\right)^{-i\kappa}
\,\sum_{i=0}^\infty b_i\left(1-\frac{r_h}{r}\right)^i.
\end{equation}
For $D>9$, some of the singularities appear in the unit circle and
one has to continue the Frobenius series through some midpoints. This
technique was recently developed in \cite{Rostworowski:2006bp}.

Substituting the above series into Eq. (\ref{radial}), one can
obtain a $(2D-5)$-term relation for the coefficients $b_i$
\begin{equation}
\sum_{j=0}^{min(2D-6,i)} c_{j,i}^{(2D-5)}(\omega,\lambda)\,b_{i-j}=0,\quad
{\rm for}\,\,i>0\,.
\end{equation}
Each of the functions $c_{j,i}^{(2D-5)}(\omega,\lambda)$ has an analytical
form in terms of $s$, $i$, $\omega$ and $\lambda$, for a specific
dimensionality $D$ and each integer $0\leq j\leq 2D-6$.

We now decrease by one the number of terms in the relation
\begin{equation}\label{srcRE}
\sum_{j=0}^{min(k,i)}c_{j,i}^{(k+1)}(\omega,\lambda)\,b_{i-j}=0\,,
\end{equation}
i.e. find the $c_{j,i}^{(k)}(\omega,\lambda)$ that satisfy
the equation
\begin{equation}\label{finRE}
\sum_{j=0}^{min(k-1,i)}c_{j,i}^{(k)}(\omega,\lambda)\,b_{i-j}=0\,.
\end{equation}
For $i\geq k$, we can rewrite the above as
\begin{equation}\label{subsRE}
\frac{c_{k,i}^{(k+1)}(\omega,\lambda)}{c_{k-1,i-1}^{(k)}(\omega,\lambda)}
\sum_{j=1}^{k}c_{j-1,i-1}^{(k)}(\omega,\lambda)\,b_{i-j}=0.
\end{equation}
Subtracting (\ref{subsRE}) from (\ref{srcRE}) we find the relation
(\ref{finRE}) explicitly, thus we obtain:
\begin{eqnarray}
&&c_{j,i}^{(k)}(\omega,\lambda) = c_{j,i}^{(k+1)}(\omega,\lambda),\qquad
{\rm for}\,\,j=0,\,\,\mbox{or}\,\,i<k,\nonumber \\[2mm]
&&c_{j,i}^{(k)}(\omega,\lambda) =
c_{j,i}^{(k+1)}(\omega,\lambda)-\frac{c_{k,i}^{(k+1)}(\omega,\lambda)\,
c_{j-1,i-1}^{(k)}(\omega,\lambda)}{c_{k-1,i-1}^{(k)}(\omega,\lambda)}\,,
\nonumber\label{GaussElimination}
\end{eqnarray}
otherwise.
This procedure is called \emph{Gaussian eliminations}, and allows us
to determine numerically the coefficients in the three-term relation
\begin{eqnarray}
&&c_{0,i}^{(3)}\,b_i+c_{1,i}^{(3)}\,b_{i-1}+c_{2,i}^{(3)}\,b_{i-2}=0, \quad
{\rm for}\,\,i>1\nonumber\\[1mm]
&&c_{0,1}^{(3)}\,b_1+c_{1,1}^{(3)}\,b_0=0,
\end{eqnarray}
for given $\omega$ (and therefore $\lambda$) up to any finite $i$. The
complexity of the procedure is \emph{linear} with respect to $i$ and $k$.

The requirement that the Frobenius series be convergent at spatial
infinity implies that
\begin{equation}
0=c_{1,1}^{(3)}-\frac{c_{0,1}^{(3)}c_{2,2}^{(3)}}{c_{1,2}^{(3)}-}
\,\frac{c_{0,2}^{(3)}c_{2,3}^{(3)}}{c_{1,3}^{(3)}-}\ldots\,.
\end{equation}
The above can be inverted $N$ times to give
\begin{eqnarray}
&~&
\hspace*{-0.5cm}c_{1,N+1}^{(3)}-\frac{c_{2,N}^{(3)}c_{0,N-1}^{(3)}}{c_{1,N-1}^{(3)}-}
\,\frac{c_{2,N-1}^{(3)}c_{0,N-2}^{(3)}}{c_{1,N-2}^{(3)}-}\ldots\,
\frac{c_{2,2}^{(3)}c_{0,1}^{(3)}}{c_{1,1}^{(3)}}\nonumber\\
&~&=\frac{c_{0,N+1}^{(3)}c_{2,N+2}^{(3)}}{c_{1,N+2}^{(3)}-}
\frac{c_{0,N+2}^{(3)}c_{2,N+3}^{(3)}}{c_{1,N+3}^{(3)}-}\ldots\,.
\end{eqnarray}
This equation with \emph{infinite continued fraction} on the
right-hand side can be solved numerically by minimizing the
absolute value of the difference between the left- and right-hand
sides. The equation has an infinite number of roots (corresponding
to the QN spectrum), but for each $N$ the most stable root is
different. In general, we have to use the $N$ times inverted
equation to find the $N$-th QN mode. The requirement that the
continued fraction be itself convergent allows us to limit its
depth by some large value, always ensuring that an increase in
this value does not change the final results within the desired
precision.  To check the correctness of the obtained results we
compared them with those obtained in \cite{KK} with the help of
WKB method \cite{will}. The WKB method is known to give very
accurate results for higher values of $\ell$ \cite{WKBaccuracy}.

The complete numerical code for determining the
QN spectrum was written in MATHEMATICA\textregistered{}, and is
available from the last author upon request.


\section{Numerical results}

We now proceed to the presentation of our numerical results for the QN
spectra of brane-localised fields with spin $s=0,1/2$ and 1 living on
a higher-dimensional Kerr black-hole induced background. In what follows,
we discuss separately each species of fields and the dependence of its
QN spectrum on the angular momentum of the black hole for arbitrary
dimensionality of spacetime.

\subsection{Brane-Localised Scalar Fields}

Starting with the case of brane fields with spin $s=0$, in Table I we display
the fundamental modes ($n=0$) of the QN spectrum for the partial mode
$(\ell=0, m=0)$. Throughout this paper, unless explicitly stated, we
will be focusing on the fundamental modes of the spectrum as they are
found to dominate the signal, with the contribution of the second mode
coming out to be about 1\%, and that of the higher overtones even smaller.
In Table I, the quasi-normal frequencies for the scalar field are
presented in terms of the angular momentum of the black hole,
measured in units of the black hole horizon radius $r_h$, and for the
values $D=5,6$ and 7 of the dimensionality of spacetime --  note that an
upper bound exists on the value of the angular momentum parameter of the
black hole given by $a_*^\text{max}=(D-2)/2$ \cite{harris} and
following from the assumption that the black hole was created by the collision
of two particles with impact parameter $b < r_h$; nevertheless, for
completeness, here we extend our analysis to higher values of $a_*$
thus allowing for black holes created by alternative mechanisms.
From the entries of Table I,
one easily observes that, as $a_*$ increases, both the real and imaginary
part of the QN frequency decreases. The latter feature makes the damping
time longer and the field oscillations on the brane longer-lived. On the
other hand, for fixed $a_*$ and variable $D$, we find the same behaviour
as in the case of a non-rotating black hole \cite{KK} with the real part
of the QN frequency being suppressed and the imaginary part enhanced, as
$D$ increases.

\begin{table}
\caption{QN frequencies $\omega r_h$ for scalar fields, for the mode $\ell=0$, $m=0$
and various values of the BH angular momentum $a_*\equiv a/r_h$ and bulk
dimensionality $D$}
\smallskip
\begin{ruledtabular}
\begin{tabular}{cccc}
  \hspace*{0.1cm} $a_*$ \hspace*{0.1cm} &\hspace*{0.4cm} $D=5$
  \hspace*{0.4cm} & \hspace*{0.4cm} $D=6$ \hspace*{0.4cm} &
  \hspace*{0.4cm} $D=7$ \hspace*{0.4cm}\\
  \hline\\[-2mm]
  0.0&0.27339-0.41091i&0.21040-0.57167i&0.07281-0.61990i\\
  0.5&0.24918-0.35635i&0.18233-0.51895i&0.04666-0.54481i\\
  1.0&0.19904-0.26464i&0.12312-0.42301i&0.01619-0.40662i\\
  1.5&0.15277-0.19716i&0.06734-0.33503i& -- \\
  2.0&0.11874-0.15493i&0.03337-0.26202i& -- \\
  2.5&0.09556-0.12834i&0.01790-0.20889i& -- \\
  3.0&0.08032-0.11010i&0.01081-0.17195i& -- \\
  3.5&0.06978-0.09604i& -- & -- \\
  4.0&0.06167-0.08478i& -- & -- \\
  4.5&0.05513-0.07574i& -- & -- \\
  5.0&0.04978-0.06843i& --& -- \\
  5.5&0.04536-0.06241i& -- & -- \\
  6.0&0.04166-0.05736i& -- & -- \\
\end{tabular}
\end{ruledtabular}
\end{table}

As the reader notes, the study of the mode $(\ell=0, m=0)$ is incomplete;
this is due to the very slow convergence of the Frobenius procedure for this
particular mode as $D$ and $a$ become large. Although upon allowing sufficient
computing time the missing quasi-normal modes could be found, the
timescale for doing so does come out to be unrealistically large. For
this reason, in Table II we present detailed numerical results for the
next partial mode $(\ell=1, m=0)$. These results confirm the monotonic
suppression of both the real and imaginary part of the quasinormal frequency,
as $a_*$ increases, thus pointing to the conclusion that this behaviour is
independent of the particular multipole number. In addition, the more complete
display of results available for this mode reveal some additional features
of the spectrum. To start with, the increase in $\omega_\text{Re}$ found
in \cite{KK} for the mode $(\ell=1, m=0)$ as $D$ goes from 5 to 6  is actually
followed by a decrease as $D$ increases further, with this behaviour persisting
for all values of $a_*$; on the other hand, the behaviour of $\omega_\text{Im}$
continues to be monotonically increasing as $D$ takes larger values.

\begin{figure}
\begin{center}
\includegraphics[height=5.4cm, clip]{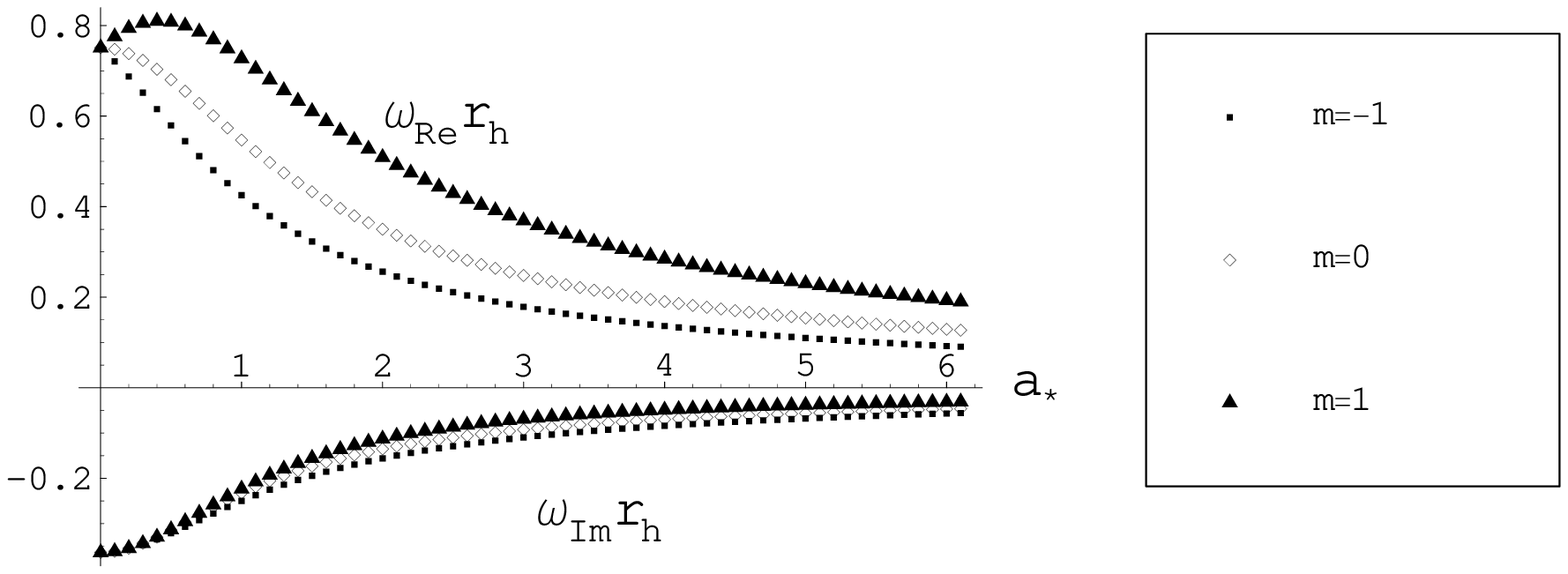}
\caption{Fundamental QN frequencies of a scalar field, as a function of $a_*$,
for the multipole mode $\ell=1$, spacetime dimensionality $D=5$,
and $m=1,0,-1$ (depicted by triangles, rhombuses and squares,
respectively).}
\label{fig-scalars}
\end{center}
\end{figure}

\begin{figure}[ht]
\begin{center}
\includegraphics[height=5.2cm, clip]{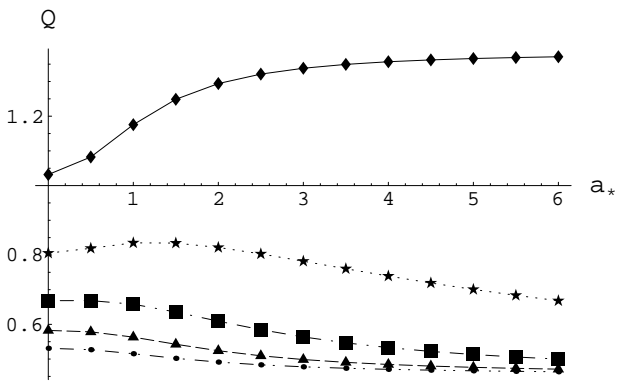}
\caption{Quality factor for the scalar field multipole mode ($\ell=1$, $m=0$),
and for $D = 5, 6,...9$ (from top to bottom).} \label{Q-sc}
\end{center}
\end{figure}

One additional important feature emerging from the entries of Table II is the
existence of asymptotic values for both $\omega_\text{Re}$ and $\omega_\text{Im}$,
as $a_* \rightarrow \infty$. These asymptotic values appear for all values of $D$,
and for all scalar modes. In Fig. \ref{fig-scalars}, we display the QN spectra
for the scalar modes $(\ell=1$, $m=-1,0,1$) as a function of the black hole
angular momentum parameter $a_*$, for fixed dimensionality ($D=5$). The presence
of asymptotic values for both $\omega_\text{Re}$ and $\omega_\text{Im}$ for all
modes is obvious. Coming back to the entries of Table II, one may observe the
potential existence of an asymptotic value for the imaginary part of the QN
frequency also as $D$ adopts large values: although $\omega_\text{Im}$ clearly
follows an increasing pattern, the step becomes gradually smaller as $D$ increases
-- we will return to this point later on.

Finally, a point to be addressed is the quality factor of the black hole as an
oscillator, defined by
\begin{equation}
Q \sim \frac{1}{2} \frac{\omega_\text{Re}} {|\omega_\text{Im}|}\,.
\label{quality}
\end{equation}
The larger the above factor is, the better oscillator the corresponding field
is. We find that the quality factor for scalar fields has a distinctly
different behavior for different values of $D$: in the case of $D=5$ there
is a monotonic increase of the quality factor $Q$, as a function of the
rotation parameter $a$, that renders the scalar fields better oscillators
as the black hole rotates faster. However, as $D$ increases, the increase
in $Q$ is soon followed by a suppression, and, for large enough values of
$D$, $Q$ is monotonically decreasing. This behavior is depicted in Fig. \ref{Q-sc}
for the scalar mode $(\ell=1, m=0$). Although, in principle, the exact value
of the quality factor is determined by all parameters of the black hole
($M_\text{BH}$, $a$) and of the propagating field ($s$, $\ell$, $m$, $n$),
it seems that, in the case of scalar fields and of the lower multipole
modes, there are two main rival factors: the angular momentum of the black
hole, that tends to increase $Q$, and the spacetime dimensionality, that
tends to decrease it. As is evident from Fig. \ref{Q-sc}, in all cases, the
quality factor reaches an asymptotic constant value as $a_* \rightarrow \infty$,
in agreement with the discussion above.


\begin{table*}
\caption{QN frequencies $\omega r_h$ for scalar fields, for the mode $\ell=1$, $m=0$
and various values of the BH angular momentum $a_*\equiv a/r_h$ and bulk
dimensionality $D$}
\smallskip
\begin{ruledtabular}
\begin{tabular}{cccccc}
\hspace*{0.1cm} $a_*$ \hspace*{0.1cm} & \hspace*{0.1cm} $D=5$ \hspace*{0.1cm}
& \hspace*{0.1cm} $D=6$ \hspace*{0.1cm} & \hspace*{0.1cm} $D=7$ \hspace*{0.1cm}
& \hspace*{0.1cm} $D=8$ \hspace*{0.1cm} & \hspace*{0.1cm} $D=9$ \hspace*{0.1cm}\\
  \hline\\[-2mm]
  0.0&0.75085-0.36387i&0.81265-0.50454i&0.81678-0.61105i&0.79233-0.68002i&0.76177-0.71730i\\
  0.5&0.68028-0.31425i&0.74928-0.45732i&0.75414-0.56410i&0.72911-0.63029i&0.69952-0.66372i\\
  1.0&0.54711-0.23270i&0.62853-0.37640i&0.63235-0.48047i&0.60568-0.53769i&0.57925-0.56198i\\
  1.5&0.43282-0.17331i&0.52186-0.31276i&0.52101-0.40993i&0.49343-0.45466i&0.47184-0.46989i\\
  2.0&0.35010-0.13527i&0.44102-0.26836i&0.43401-0.35623i&0.40757-0.38891i&0.39096-0.39791i\\
  2.5&0.29112-0.11016i&0.38023-0.23672i&0.36761-0.31440i&0.34386-0.33751i&0.33138-0.34279i\\
  3.0&0.24803-0.09267i&0.33336-0.21315i&0.31655-0.28076i&0.29609-0.29694i&0.28670-0.30011i\\
  3.5&0.21554-0.07987i&0.29619-0.19478i&0.27671-0.25307i&0.25944-0.26447i&0.25230-0.26642i\\
  4.0&0.19033-0.07013i&0.26599-0.17991i&0.24510-0.22993i&0.23065-0.23807i&0.22512-0.23930i\\
  4.5&0.17025-0.06249i&0.24096-0.16751i&0.21961-0.21035i&0.20751-0.21627i&0.20317-0.21707i\\
  5.0&0.15392-0.05633i&0.21988-0.15690i&0.19873-0.19364i&0.18856-0.19802i&0.18508-0.19854i\\
  5.5&0.14041-0.05128i&0.20190-0.14766i&0.18136-0.17925i&0.17276-0.18254i&0.16994-0.18289i\\
  6.0&0.12904-0.04705i&0.18640-0.13950i&0.16672-0.16675i&0.15940-0.16926i&0.15708-0.16951i\\
\end{tabular}
\end{ruledtabular}
\end{table*}

\subsection{Brane-Localised Gauge Bosons}

We now turn to the case of gauge fields propagating in the induced-on-the-brane
background of a higher-dimensional rotating black hole. Our results are displayed
in Table III, and similarly to the case of scalar fields, they reveal the
monotonic suppression of both the real and imaginary part of the QN frequency,
as the angular momentum of the black hole increases. In addition, as
$a_* \rightarrow \infty$, constant asymptotic values emerge for both
$\omega_\text{Re}$ and $\omega_\text{Im}$ of the quasinormal frequency.
This is obvious from the results displayed in Table III for the mode
$(\ell=1, m=0)$, but it has been found to hold also for the modes
$m=\pm 1$.

When $a_*$ is kept fixed while $D$ varies, $\omega_\text{Re}$ first increases
as $D$ goes from 5 to 6, in accordance with the generic behaviour found in
\cite{KK}, but as $D$ increases further, a monotonic suppression takes over
again. Also in agreement with the non-rotating case \cite{KK}, the imaginary
part of the QN frequency monotonically increases, as $D$ grows, making the
gauge field perturbations shorter-lived. The existence of a potential
asymptotic value for $\omega_\text{Im}$, as $D \rightarrow \infty$, can
again be inferred from the entries of Table III.

\begin{figure}[floatfix]
\begin{center}
\includegraphics[height=5cm, clip]{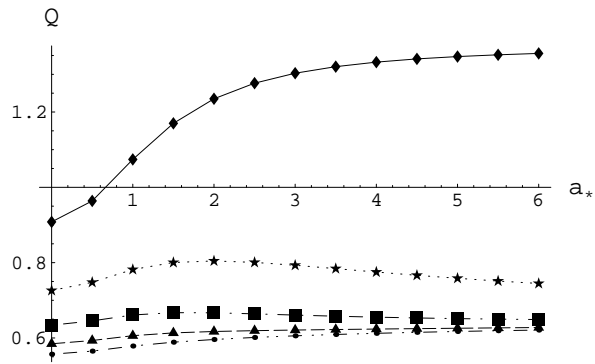}
\caption{Quality factor for the gauge field multipole mode ($\ell=1$, $m=0$),
and for $D = 5, 6,...9$ (from top to bottom).} \label{Q-gb}
\end{center}
\end{figure}

The quality factor $Q$ for gauge bosons is again strongly dependent on the
value of the dimensionality of spacetime $D$. Its behavior for the mode
$(\ell=1, m=0)$ and for the values $D=5,6,...,9$ is shown in Fig. \ref{Q-gb}.
As is obvious, for $D=5$ we obtain again a sharp monotonic increase, as
$a_*$ increases. For $D=6,7$, we find, as in the case of scalar fields,
a non-monotonic behavior with an initial increase followed by a decrease.
The deviation from the case of scalar fields arises for $D>7$, where the
monotonic suppression found in that case gives its place to a monotonic
enhancement, although a marginal one.


\begin{table*}
\caption{QN frequencies $\omega r_h$ for gauge fields, for the mode $\ell=1$, $m=0$
and various values of the BH angular momentum $a_*\equiv a/r_h$ and bulk
dimensionality $D$}
\begin{ruledtabular}
\begin{tabular}{cccccc}
 \hspace*{0.1cm} $a_*$ \hspace*{0.1cm} & \hspace*{0.1cm} $D=5$ \hspace*{0.1cm}
& \hspace*{0.1cm} $D=6$ \hspace*{0.1cm} & \hspace*{0.1cm} $D=7$ \hspace*{0.1cm}
& \hspace*{0.1cm} $D=8$ \hspace*{0.1cm} & \hspace*{0.1cm} $D=9$ \hspace*{0.1cm}\\
  \hline\\[-2mm]
  0.0&0.57667-0.31749i&0.58397-0.40214i&0.56925-0.44900i&0.55266-0.47270i&0.53963-0.48451i\\
  0.5&0.52746-0.27360i&0.54068-0.36152i&0.52762-0.40874i&0.51222-0.43168i&0.50037-0.44273i\\
  1.0&0.43057-0.20043i&0.45531-0.29121i&0.44497-0.33654i&0.43206-0.35645i&0.42276-0.36532i\\
  1.5&0.34379-0.14693i&0.37737-0.23574i&0.36855-0.27619i&0.35810-0.29186i&0.35125-0.29827i\\
  2.0&0.27931-0.11307i&0.31725-0.19716i&0.30905-0.23173i&0.30067-0.24357i&0.29569-0.24807i\\
  2.5&0.23272-0.09115i&0.27183-0.16975i&0.26399-0.19877i&0.25728-0.20768i&0.25363-0.21086i\\
  3.0&0.19845-0.07615i&0.23694-0.14938i&0.22949-0.17365i&0.22410-0.18042i&0.22137-0.18273i\\
  3.5&0.17253-0.06533i&0.20952-0.13360i&0.20255-0.15396i&0.19817-0.15919i&0.19609-0.16091i\\
  4.0&0.15238-0.05718i&0.18751-0.12096i&0.18105-0.13815i&0.17746-0.14227i&0.17584-0.14358i\\
  4.5&0.13633-0.05083i&0.16950-0.11058i&0.16356-0.12521i&0.16058-0.12851i&0.15931-0.12952i\\
  5.0&0.12327-0.04575i&0.15453-0.10187i&0.14908-0.11443i&0.14659-0.11712i&0.14557-0.11792i\\
  5.5&0.11245-0.04159i&0.14190-0.09445i&0.13693-0.10532i&0.13482-0.10754i&0.13399-0.10818i\\
  6.0&0.10335-0.03812i&0.13113-0.08804i&0.12658-0.09753i&0.12478-0.09939i&0.12409-0.09991i\\
\end{tabular}
\end{ruledtabular}
\end{table*}

\subsection{Brane-Localised Fermions}

We finally address the case of brane-localised spinor fields. In Tables
IV and V, we present results for the two lowest multipole modes
$(\ell=1/2, m=1/2)$ and $(\ell=1/2, m=-1/2)$, respectively, for various
values of the black hole angular momentum parameter and dimensionality
of spacetime. Unlike the previously studied cases of scalars and gauge
fields, a deviation is observed in the behavior of these two modes,
therefore we discuss both of them here separately.

Starting from the mode $(\ell=1/2, m=1/2)$, whose QN spectrum is presented
in Table IV, we find that, in terms of $a_*$, the generic behavior noticed
for the other two species of fields also prevails here: both
$\omega_\text{Re}$ and $\omega_\text{Im}$ monotonically decrease as
the angular momentum of the black hole increases, eventually reaching
constant asymptotic values. On the other hand, for fixed $a_*$ and
variable $D$, $\omega_\text{Im}$ increases monotonically, as for the
other two species of fields, with the only difference being the
absence of an asymptotic value as $D \rightarrow \infty$. A further
deviation arises in the dependence of $\omega_\text{Re}$ on $D$ with
its exact behavior depending on the value of the black hole angular
momentum: for low values of $a_*$, $\omega_\text{Re}$ follows the traditional
increase-followed-by-a-decrease pattern, while for high values of $a_*$
a monotonic decrease for all values of $D$ is observed instead.


\begin{table*}[t]
\caption{QN frequencies $\omega r_h$ for spinor fields, for the mode $\ell=1/2$, $m=1/2$
and various values of the BH angular momentum $a_*\equiv a/r_h$ and bulk
dimensionality $D$}
\begin{ruledtabular}
\begin{tabular}{cccccc}
  \hspace*{0.1cm} $a_*$ \hspace*{0.1cm} & \hspace*{0.1cm} $D=5$ \hspace*{0.1cm}
& \hspace*{0.1cm} $D=6$ \hspace*{0.1cm} & \hspace*{0.1cm} $D=7$ \hspace*{0.1cm}
& \hspace*{0.1cm} $D=8$ \hspace*{0.1cm} & \hspace*{0.1cm} $D=9$ \hspace*{0.1cm}\\
  \hline\\[-2mm]
  0.0&0.44130-0.35984i&0.45332-0.51049i&0.41925-0.65204i&0.33208-0.78624i&0.31438-1.28297i\\
  0.5&0.45308-0.29340i&0.45965-0.43444i&0.41755-0.55905i&0.32135-0.65497i&0.30694-1.11216i\\
  1.0&0.39498-0.19854i&0.40166-0.33780i&0.34902-0.44834i&0.25178-0.49810i&0.23566-0.95459i\\
  1.5&0.32503-0.13271i&0.33293-0.27005i&0.27090-0.36141i&0.19483-0.37575i&0.18243-0.85971i\\
  2.0&0.26747-0.09332i&0.27517-0.22679i&0.20951-0.29541i&0.15719-0.29478i&0.14483-0.80778i\\
  2.5&0.22380-0.06941i&0.22978-0.19751i&0.16694-0.24551i&0.13128-0.24071i&0.11788-0.77704i\\
  3.0&0.19081-0.05414i&0.19420-0.17589i&0.13772-0.20815i&0.11251-0.20284i&0.09839-0.75720i\\
  3.5&0.16548-0.04387i&0.16604-0.15868i&0.11695-0.17989i&0.09836-0.17507i&0.08409-0.74360i\\
  4.0&0.14561-0.03665i&0.14356-0.14430i&0.10155-0.15807i&0.08734-0.15390i&0.07332-0.73388i\\
  4.5&0.12969-0.03137i&0.12551-0.13196i&0.08972-0.14082i&0.07852-0.13726i&0.06498-0.72671i\\
  5.0&0.11671-0.02740i&0.11091-0.12125i&0.08036-0.12689i&0.07131-0.12386i&0.05835-0.72126i\\
  5.5&0.10594-0.02433i&0.09900-0.11190i&0.07277-0.11543i&0.06531-0.11283i&0.05296-0.71701i\\
  6.0&0.09688-0.02192i&0.08918-0.10371i&0.06649-0.10585i&0.06024-0.10360i&0.04849-0.71363i\\
\end{tabular}
\end{ruledtabular}
\end{table*}

\begin{table*}
\caption{QN frequencies $\omega r_h$ for spinor fields, for the mode $\ell=1/2$,
$m=-1/2$ and various values of the BH angular momentum $a_*\equiv a/r_h$ and bulk
dimensionality $D$}
\begin{ruledtabular}
\begin{tabular}{cccccc}
  \hspace*{0.1cm} $a_*$ \hspace*{0.1cm} & \hspace*{0.1cm} $D=5$ \hspace*{0.1cm}
& \hspace*{0.1cm} $D=6$ \hspace*{0.1cm} & \hspace*{0.1cm} $D=7$ \hspace*{0.1cm}
& \hspace*{0.1cm} $D=8$ \hspace*{0.1cm} & \hspace*{0.1cm} $D=9$ \hspace*{0.1cm}\\
  \hline\\[-2mm]
  0.0&0.44130-0.35984i&0.45332-0.51049i&0.41925-0.65204i&0.33208-0.78624i&0.31438-1.28297i\\
  0.5&0.35706-0.32705i&0.37986-0.49522i&0.34876-0.66241i&0.25678-0.84279i&0.17791-1.40707i\\
  1.0&0.26937-0.25567i&0.30417-0.43439i&0.27267-0.63077i&0.17717-0.87878i&0.04019-1.48580i\\
  1.5&0.20663-0.19865i&0.25106-0.37925i&0.21963-0.60756i&0.14808-0.92662i& -- \\
  2.0&0.16466-0.15934i&0.21607-0.33850i&0.18960-0.60173i&0.16145-0.97336i& -- \\
  2.5&0.13587-0.13196i&0.19246-0.30939i&0.17898-0.60691i& -- & -- \\
  3.0&0.11525-0.11217i&0.17585-0.28853i&0.18277-0.61502i& -- & -- \\
  3.5&0.09988-0.09735i&0.16372-0.27344i&0.19531-0.62133i& -- & -- \\
  4.0&0.08804-0.08588i&0.15462-0.26239i&0.21210-0.62451i& -- & -- \\
  4.5&0.07865-0.07678i&0.14768-0.25420i&0.23044-0.62496i& -- & -- \\
  5.0&0.07105-0.06939i&0.14236-0.24804i&0.24898-0.62352i& -- & -- \\
  5.5&0.06477-0.06327i&0.13830-0.24332i& -- & -- & -- \\
  6.0&0.05949-0.05814i&0.13527-0.23962i& -- & -- & -- \\
\end{tabular}\medskip
\end{ruledtabular}
\end{table*}

Turning to the mode $(\ell=1/2, m=-1/2)$ and to the entries of Table V,
the deviation in the behavior
of its QN spectrum is even more striking. As $a_*$ increases, its
$\omega_\text{Re}$ and $\omega_\text{Im}$ monotonically decrease towards
some constant asymptotic values, as expected, only for the values $D=(5,6)$.
As $D$ increases further, the initial phase of decrease for $\omega_\text{Re}$
follows a phase of enhancement; $\omega_\text{Im}$ behaves even
more singularly by oscillating for $D=7$ and monotonically increasing
for $D \geq 8$. When $a_*$ is kept fixed and $D$ varies, the previously
observed behavior is followed only for low values of $a_*$, with
$\omega_\text{Re}$ increasing until $D=6$ and decreasing for greater
values of $D$; for values of the angular momentum parameter higher than
$a_*=3$, the enhancement phase is extended at least up to $D=7$. Finally,
$\omega_\text{Im}$ follows the traditional increasing pattern
with $D$, however, as in the case of $m=1/2$, there is no sign for
the existence of an asymptotic value as $D \rightarrow \infty$.

The alert reader might have noticed that the deviations in the
behavior of the QN spectrum for the aforementioned spinor modes arise
in the case of either large $a_*$ or large $D$. In section IV.A.
we commented on the very slow convergence of the Frobenius procedure,
for exactly these ranges of parameters, that failed to give us
complete results for the lowest scalar mode $(\ell=0, m=0)$. The
same problem appeared also in the case of the spinor mode
$(\ell=1/2, m=-1/2$), as it is obvious from Table V. Although
numerical results were successfully derived for smaller values
of either $a_*$ and/or $D$, it might be possible that the looming bad
convergence behavior of our numerical method contaminated these
results, too. Another reason for the observed modifications might
be the fact that the equations for the QN modes corresponding to
different $\ell$'s are entangled and, at times, it is extremely
difficult to distinguish QN modes by their multipole number,
especially in the case of ultra-spinning black holes. For the
above reasons, the determined results for spinor fields for high
values of either $a_*$ or $D$ should be taken with caution.

\begin{figure}
\begin{center}
\includegraphics[height=5.3cm, clip]{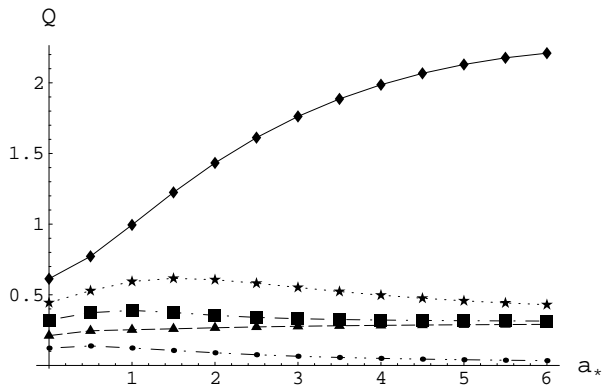}
\caption{Quality factor for the spinor field multipole mode ($\ell=1/2$, $m=1/2$),
and for $D = 5, 6,...9$ (from top to bottom).} \label{Q-fm1}
\end{center}
\end{figure}
\begin{figure}
\begin{center}
\includegraphics[height=5.1cm, clip]{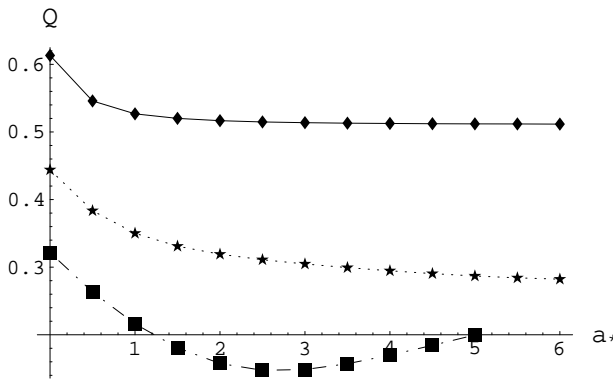}
\caption{Quality factor for the spinor field multipole mode ($\ell=1/2$, $m=-1/2$),
and for $D = 5, 6, 7$ (from top to bottom).} \label{Q-fm2}
\end{center}
\end{figure}

Finally, the quality factor for the spinor field mode ($\ell=1/2,
m=1/2$) closely follows the behavior observed in the case of
scalar and gauge fields, as it is obvious from Fig. \ref{Q-fm1}.
For $D=5$, $Q$ has a monotonically increasing dependence on $a_*$,
that brings it to a value higher than for other species, thus
rendering the fermions the best oscillators among the lowest
multipole modes. For $D \geq 6$, the quality factor reaches a
maximum value and then follows a decreasing pattern; for large
values of $D$, the change in the value of $Q$, as $a_*$ varies,
is only marginal. Turning to the mode ($\ell=1/2, m=-1/2$), we
may easily see from Fig. \ref{Q-fm2} that the quality factor
follows a radically different behavior. Although the suppression
with the dimensionality of spacetime is still present, the
enhancement in terms of $a_*$ has disappeared. For $D=(5,6)$,
we can observe a monotonic decrease in $Q$, as $a_*$ grows, while
for $D=7$ the decrease in $Q$ changes to an increase, after reaching
a minimum at $a_{*} \approx 2.5$. The absence of robust results
for this mode for $D>7$ unfortunately does not allow us to check
the dependence of its quality factor on $a_*$ for larger values of $D$.
Nevertheless, a clear conclusion can be safely drawn when the two lowest
multipole spinor modes are compared: the "co-rotating" modes with $m=1/2$
are much better oscillators than the "contra-rotating" ones with
$m=-1/2$.

\section{Conclusions}

In this paper, we have performed a comprehensive study of the fundamental
quasinormal modes of all Standard Model fields propagating on a brane
embedded in a higher-dimensional rotating black hole spacetime. The starting
point of our analysis was the equations of motion for fields with spin
$s=0,1/2$ and 1 propagating on the induced-on-the-brane background, and
their re-writing to the form of a Schr\"odinger-like, wave equation.
By using numerical techniques, based on the continued fraction method,
the quasinormal spectra of all species of fields were derived as a
function of the black hole angular momentum and dimensionality of spacetime.

\begin{figure}[b]
\begin{center}
\includegraphics[height=5.3cm, clip]{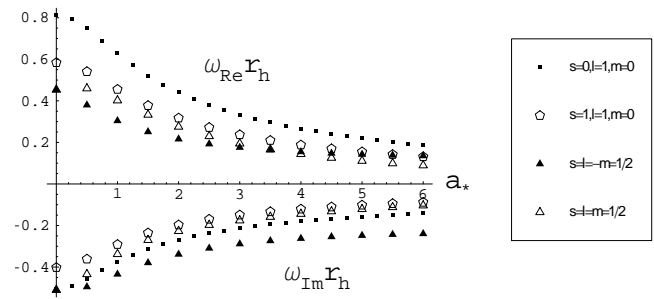}
\caption{Fundamental QN modes of fields with spin $s=0, 1/2$ and 1, for $D=6$.}
\label{D6-coll}
\end{center}
\end{figure}

The dependence of the QN spectra on the angular momentum of the black
hole was found to be universal with both the real and (the absolute value
of the) imaginary part of
the QN frequency being suppressed as $a_*$ increases. As a result,
brane-localised fields propagating on the induced 4-dimensional background
are longer-lived when the higher-dimensional black hole rotates faster.
In addition, both $\omega_\text{Re}$ and $\omega_\text{Im}$ approach
constant asymptotic values, as $a_* \rightarrow \infty$. These results
hold for all species of fields, with the only exception arising for
spinor fields with $m=-1/2$ and for $D \geq 7$. Figure \ref{D6-coll}
shows the collective behaviour of all SM fields QN spectra as a function
of $a_*$, for $D=6$, and reveals the aforementioned spin-independent
behaviour.

\begin{figure}[t]
\begin{center}
\includegraphics[height=5.3cm, clip]{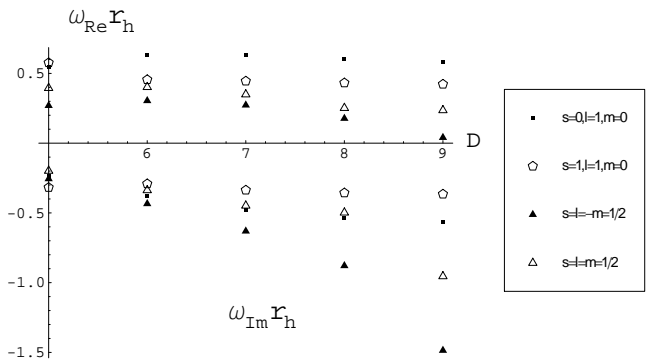}
\caption{Fundamental QN modes of fields with spin $s=0, 1/2$ and 1, for $a_{*}=1$.}
\label{a1-coll}
\end{center}
\end{figure}

In terms of the dimensionality of spacetime, our analysis confirmed the
increase in the imaginary part of the QN frequency of all SM fields,
as $D$ increases, observed in the case of a non-rotating black hole \cite{KK}.
The same behaviour holds for arbitrary angular momentum of the black
hole, a result that points to the conclusion that an increase in
the number of transverse-to-the-brane dimensions reduces the lifetime
of the field perturbations on the brane, in both spherically- and
axially-symmetric backgrounds. With the exception again of spinor fields,
$\omega_\text{Im}$ tends again to a constant asymptotic value as
$D \rightarrow \infty$. This behavior is clearly shown in Fig.
\ref{a1-coll}, where the QN spectra of all SM fields are shown for
$a_*=1$. In the same figure, we also display the collective behavior
of the real part of the QN frequency, $\omega_\text{Re}$, as a function
of $D$. This behavior is also found to be universal, with
$\omega_\text{Re}$ increasing up to $D=6$, in accordance with previously
found results \cite{KK}, but decreasing for higher values of $D$.

The quality factor $Q$, finally, that determines the best oscillator
among the different field perturbations, was found to depend both
on properties of the field, like its spin $s$ or multipole numbers
$(\ell, m$), as well as on properties of the gravitational background, such
as the dimensionality of spacetime and angular momentum of the black
hole. The quality factor of modes with either $m=0$ or $m>0$ was found
to decrease with $D$ but (predominantly) increase with $a_*$. As a
result, for all of these modes, $Q$ takes its largest possible value
for $D=5$ and large $a_*$ - among the different species, the fermionic
modes were found to assume the largest value of $Q$ for these
parameter values. The situation however changes radically for modes
with $m<0$, whose quality factor is suppressed with $D$ but also
with $a_*$, rendering them worse oscillators than the ones with
$m=0$ or $m>0$.

\begin{figure}[ht]
\begin{center}
\includegraphics[height=5.3cm,clip]{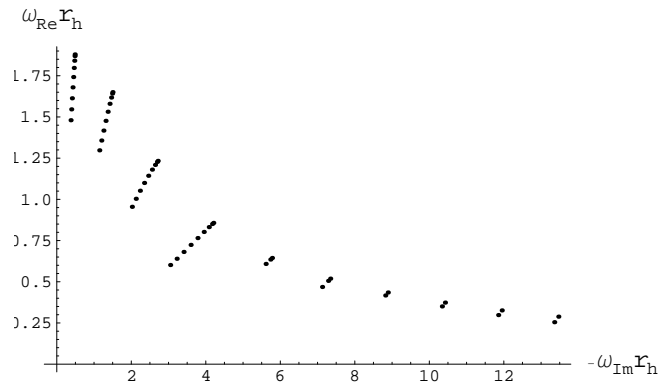}
\caption{A few higher overtones for the electromagnetic field mode
($\ell=3$, $m=0$), and $D=6$.}
\label{higher}
\end{center}
\end{figure}

Being restricted by the standard model we did not compute
quasi-normal modes of massive fields, yet, as was shown in
\cite{massivescalar} for massive scalar and in \cite{massivevector} for
massive vector fields, the QN spectrum can show rather exotic
behavior, for instance, infinitely long living modes. Also,
throughout this paper, we have restricted our study to the
fundamental overtones ($n=0$) of the QN spectra due to their
dominance in the signal. Nevertheless, higher overtones of all SM
fields with spin $s=0, 1/2$ and 1 were also looked at in the
context of our analysis. Their QN spectra were derived for the
range $(0,r_h)$ of the angular momentum parameter $a$ of the black
hole with a step of $r_h/8$. In Fig.
\ref{higher}, we display a few of the higher overtones for a
brane-localised gauge field with $(\ell=3, m=0$). For all higher
overtones, it was found that an increase in $a_*$ leads again to
the decrease of the absolute values of both the real and imaginary
parts of their QN frequencies.

We would like to finish our work with some important remarks. During our
analysis, we made two simplifying assumptions. Firstly, our study was
restricted to the case of higher-dimensional black holes with only one
non-zero rotating parameter; that was due to the assumption that the
colliding particles, that give rise to the black hole, were restricted
to live on an infinitely-thin brane, and therefore had a non-vanishing
impact parameter only along an axis parallel to the brane. In a more
realistic case, where the brane is assumed to have a finite thickness
-- generically of the order of the fundamental length --
the colliding particles can have non-zero impact parameters along the
bulk directions, and thus give rise to a black hole with more than one
rotating parameters. In more sophisticated brane-world models, that
attack phenomenological problems like fast proton decay, large
neutron-antineutron oscillations or FCNC, the quarks may be located
at different 3-dimensional slices along the bulk and thus have,
by construction, large separations along
the bulk directions \cite{Dai}. It seems therefore imperative
that a more comprehensive study of higher-dimensional black holes, with
all possible components of the angular momentum taken into account, need
to be done in the future (for scalar particles in $D=5$, the general
equation of motion has been derived and shown to be separable in
\cite{FS}).
The second assumption made in our analysis was that the brane tension
was much smaller than the mass of the produced black hole and thus it
could be safely ignored. However, the "friction" between the black hole
and the brane can alter the components of the angular momentum of the
former \cite{FFS}, and excite brane degrees of freedom that could be
interpreted as quasi-normal modes. The analysis of such a realistic
set-up, that would study the change in the spectrum of the quasinormal
frequencies found here due to the brane-black hole interaction, could also
be the objective of a future work.


\begin{acknowledgments}

P.K. acknowledges financial support from the U.K. Particle Physics and Astronomy
Research Council (Grant Number PPA/A/S/2002/00350).
The work of R. K. and A. Z. was supported by \emph{Funda\c{c}\~{a}o de Amparo
\`{a} Pesquisa do Estado de S\~{a}o Paulo (FAPESP)}, Brazil.

\end{acknowledgments}


%
\end{document}